\documentstyle[11pt,aaspp4,amssym]{article}
\lefthead{Sofia et al.}
\righthead{Carbon Toward HD 24534}
\begin{document}
\title{A Reanalysis of the Carbon Abundance in the Translucent Cloud Toward 
HD 24534  \footnotemark[1]}

\author{Ulysses J. Sofia, Edward L. Fitzpatrick}
\affil{Department of Astronomy and Astrophysics, Villanova University,
    Villanova, PA 19085}

\and

\author{David M. Meyer}
\affil{Department of Physics and Astronomy, Northwestern University, 
Evanston, IL 60208}

\footnotetext[1] {Based on observations obtained with the NASA/ESA {\it
Hubble Space Telescope} through the Space Telescope Science Institute,
which is operated by the Association of Universities for Research in
Astronomy, Inc., under NASA contract NASA-26555.}

\begin{abstract}
We have reanalyzed the Goddard High Resolution Spectrograph data set
presented by Snow et al. which contains the interstellar intersystem 
\ion {C} {2} {\small]} $\lambda$2325 line through the translucent 
cloud toward
HD~24534 (X~Persei). In contrast to the results of Snow et al., we clearly
detect the \ion{C}{2}{\small]} feature at the 3$\sigma$ confidence level and
measure a C$^{+}$ column density of $2.7 \pm 
0.8 \times 10^{17}$ cm$^{-2}$. Accounting for the \ion{C}{1} column density
along the line of sight, we find 10$^{6}$ C/H = 106 $\pm$ 38 in the 
interstellar gas toward this star. This gas-phase carbon-to-hydrogen ratio 
suggests that slightly more carbon depletion may be occurring
in translucent as compared to diffuse clouds. The average diffuse-cloud C/H,
however, is within the 1$\sigma$ uncertainty of the measurement toward 
HD 24534. We therefore cannot rule out the possibility 
that the two cloud types have comparable gas-phase C/H, and therefore 
comparable depletions of carbon.

\end{abstract}


\section{Introduction}

Interstellar carbon is an important contributor to interstellar 
molecular chemistry and cloud cooling, and is thought to be a 
major component of interstellar dust (\markcite{DW81}Duley \& 
Williams 1981; \markcite{DRAI89}Draine 1989; \markcite{JLM92}Joblin, 
L\'eger \& Martin 1992). Unfortunately, reliable measurements of the gas-phase
interstellar carbon abundance have been elusive because the available 
absorption features from the dominant ionization stage, C$^{+}$, are either so
strong as to be hopelessly saturated ($\lambda$1334) or so weak as to be 
undetectable ($\lambda$2325). In recent years, however, this situation has 
changed and high precision measurements of the \ion{C}{2}{\small]} 
$\lambda$2325
line with the Goddard High Resolution Spectrograph (GHRS) aboard the 
{\it Hubble Space Telescope} (HST) have yielded 6 reliable 
($>$ 2$\sigma$) measurements of the gas-phase interstellar
carbon abundance in diffuse clouds (\markcite{CARD93}Cardelli et al. 1993; 
\markcite{CARD96}Cardelli et al. 1996; \markcite{SOFI97}Sofia et al. 
1997), one upper limit in diffuse clouds (\markcite{CARD96}Cardelli et 
al. 1996), and 2 upper limits in translucent clouds 
(\markcite{SNOW96}Snow et al. 1996; \markcite{SNOW98}1998). 

Absorption line measurements suggest that in diffuse interstellar clouds, 
there is little exchange of carbon between the gas and dust phases 
(\markcite{CARD96}Cardelli et al. 1996; \markcite{SOFI97}Sofia et al. 
1997), i.e., a relatively constant C depletion level is observed. 
One might expect, however, that a denser cloud region, such as a 
translucent cloud, may provide a more protective environment than diffuse 
clouds, and therefore show different depletion characteristics. A 
recently reported upper limit for gas-phase carbon in a translucent cloud 
(toward HD 24534) suggested that this was indeed 
the case, and that C was substantially more depleted in this region 
as compared to the more diffuse clouds (\markcite{SNOW98}Snow et al. 1998). 

Since carbon plays such an important role in the interstellar medium 
(ISM) and its abundance is such a difficult quantity to measure, we decided 
to reexamine the spectrum through the translucent cloud toward HD 24534. 
In \S 2 we will discuss our data calibration and reductions and in 
\S 3 our results will be discussed.

\section{Observations and Reductions}

The observations of the \ion{C}{2}{\small]} $\lambda$2325 line towards HD 
24534 are described in 
\markcite{SNOW98}Snow et al. (1998). In summary, the data consist of 40 
individual, high resolution GHRS spectra taken with the 
observing strategy FP-SPLIT=4. This means that 10 spectra were taken 
at each of 4 slightly different positions on the photocathode in order to 
reduce the effects of fixed pattern noise (FPN) on the final spectrum. We 
obtained these data from the Hubble Space Telescope Data Archive and they were
processed independently by two of us (UJS and ELF).
In both cases the GHRS-team software package, CALHRS, was used to calibrate 
the spectra (Space Telescope pipeline calibrations do not differ significantly
from CALHRS). In addition, sophisticated flat fielding algorithms were 
employed to further reduce the effects of FPN, beyond the reduction afforded 
by the FP-SPLIT oberving strategy. These algorithms are described in 
\markcite{SF93}Spitzer \& Fitzpatrick (1993), \markcite{CE94}Cardelli 
\& Ebbets (1994) and \markcite{CARD95}Cardelli (1995). Once the FPN was
removed, the data were coadded using the nominal wavelengths determined 
from the GHRS carousel grating position to align the individual spectra. 
We examined the 40 individual spectra and found no reason to exclude any 
from the final coadded spectrum. 

The two independently processed final spectra are essentially identical 
and have S/N values of approximately 170 near 2325\AA. Figure 1 shows one of
the final spectra; count rate versus heliocentric wavelength are plotted for
a 3\AA\space segment centered near the expected location of 
\ion{C}{2}{\small]}
$\lambda$2325.403. The upper and lower spectra show the results with and 
without, respectively, the application of the additional FPN correction 
algorithms. The magnitude of the FPN features is small in the region of the 
detector sampled by these data and the improvement with the additional 
correction is only marginal. It is important to note, however, the the FPN 
removal algorithm has not added any spurious features to the data.

\section{Results and Discussion}

Figure 1 shows clear evidence for the presence of an absorption feature in 
the wavelength region expected for \ion{C}{2}{\small]} $\lambda$2325.4. 
Figure 2
provides an expanded view of this region (for the FPN corrected data) with
normalized flux plotted versus heliocentric velocity (top spectrum). The 
normalization was accomplished with a second order polymonial fit
to the continuum. For comparison, 
we also show the \ion{O}{1}{\small]} $\lambda$1355 feature seen toward HD 
24534 (lower spectrum). These data have been velocity-shifted by +1.0 km/sec
which is well within the velocity uncertainty of $\sim$3.5 km/sec expected for
GHRS calibrated data (Savage, Cardelli \& Sofia 1992). This comparison 
demonstrates that the absorption feature seen in the top spectrum is closely
aligned in velocity with the \ion{O}{1} line; we thus firmly identify this 
feature as due to \ion{C}{2}{\small]} $\lambda$2325.4, in strong contrast to 
the results of Snow et al. (1998).

We measure the C$^{+}$ column density implied by the \ion {C}{2}{\small]} 
feature in two ways, from the equivalent width and from component fitting. 
The equivalent width of the feature is found to be  
W$_{\lambda}$~=~0.78 $\pm$ 0.22 m\AA\space (note that this result is more 
than twice as large as the 0.33 m\AA\space 2$\sigma$ upper limit quoted 
by Snow et al. 1998). We adopt the 
oscillator strength f = 5.80 $\times$ 10$^{-8}$ for the \ion{C}{2}{\small]}
$\lambda$2325 transition from \markcite{CARD96}Cardelli et al. (1996) who 
combined the values of 
\markcite{FANG93}Fang et al. (1993) and \markcite{LENN85}Lennon et 
al. (1985).  Cardelli et al. estimate an uncertainty of $\sim$8\% in this
value. Assuming the weak-line limit, W$_{\lambda}$ implies N(C$^{+}$) = 
2.8 $\pm 0.8 \times 10^{17}$ cm$^{-2}$. The 1$\sigma$ uncertainty 
includes the effects of 
statistical error, continuum placement, and scattered light uncertainties 
added in quadrature, but does not include the oscillator strength uncertainty.
The component fitting technique (see, e.g. \markcite{SF93}Spitzer \& 
Fitzpatrick 1993) yields a similar result of N(C$^{+}$) = 2.6 $\pm 0.8
\times 10^{17}$ 
cm$^{-2}$. Again, the quoted uncertainty does not include the error in the 
oscillator strength.

Following Snow et al. (1998) we use the \markcite{DS94}Diplas 
\& Savage (1994) neutral H column density (N(\ion {H}{1}) = 5.4 $\pm$ 
0.8 $\times 10^{20}$ cm$^{-2}$) and the \markcite{MASO76}Mason et al. (1976) 
H$_{2}$ column density (N(H$_{2}$) = 1.10 $\pm$ 0.30 $\times 10^{21}$ 
cm$^{-2}$) to get the total column density of neutral hydrogen particles
toward HD 24534 N(H) = 2.74 $\pm$ 0.61 $\times$ 
10$^{21}$ cm$^{-2}$ . Taking the N(\ion{C}{1}) $\approx$ 2 $\times 10^{16}$ 
cm$^{-2}$ from Snow et al. (1998), and the average \ion{C}{2} abundance from 
our measurements (N(\ion{C}{2}) = 2.7 $\pm$ 0.8 $\times$ 10$^{17}$ cm$^{-2}$), 
we find the 
gas-phase carbon-to-hydrogen ratio is 10$^{6}$ C/H = 106 $\pm$ 38 where 
the 1$\sigma$ value takes into account the uncertainties in both the 
H and \ion{C}{2} (but not \ion{C}{1}) column densities.

The gas-phase interstellar carbon-to-hydrogen ratio in the translucent 
cloud toward HD 24534 is slightly lower than, although within the 1$\sigma$ 
error of, the average ratio seen in the diffuse clouds previously measured 
(10$^{6}$ C/H = 142 
$\pm$ 13; \markcite{SOFI97}Sofia et al. 1997). If we add the HD 24534 
sightline to the Sofia et al. sample, the average ratio becomes 10$^{6}$ 
C/H = 139 $\pm$ 12. Figure 3 shows the interstellar gas-phase C/H versus
the fraction of molecular hydrogen, f(H$_{2}$) = 
2N(H$_{2}$)/[2N(H$_{2}$) + N(\ion{H}{1})], for all of the measured diffuse 
cloud
sightlines, as well as the translucent cloud toward HD24534. This figure is 
an update 
to the Snow et al. (1998) Figure 3, except that we have only included the 
sightlines with \ion {C}{2} detections. The data points and error bars 
in Figure 3 do not agree precisely with Snow et al.'s Figure 3 for three
reasons: 1) We used a weighted average of the \markcite{BSD78}Bohlin, 
Savage \& Drake (1978) and the \markcite{DS94}Diplas \& Savage (1994) 
N(\ion{H}{1}) measurements whenever possible, 2) we have used the updated 
carbon abundance toward $\xi$ Per (\markcite{CARD96}Cardelli et al. 1996),  
and 3) we have included the uncertainties associated with the H abundances 
in our error bars.

Figure 3 shows a line plotted at the weighted average of the measured 
gas-phase C/H values (10$^{6}$ C/H = 139). This is a statistically acceptable
fit to the data.  We also plot a weighted linear fit to the C/H data in
Figure 3 (dotted line). Although the higher order term is not statistically 
justified, we show the linear fit in order to illustrate a possible trend 
in the data. Given the 
uncertainties in the measurements, it is not possible to conclude with 
certainty whether or not carbon is more depleted in the translucent cloud 
as compared to diffuse clouds. If carbon is more depleted in the 
HD 24534 cloud, it is not by the substantial amount suggested by the
\markcite{SNOW98}Snow et al. (1998) non-detection of \ion{C}{2}{\small]}
(which implied 10$^{6}$ C/H $\leq$ 44). It is 
interesting to note that C is not the only volatile element which does not 
appear to be significantly more depleted in translucent clouds versus 
diffuse clouds. \markcite{SNOW98}Snow et al. (1998) find the 
gas-phase O/H in the translucent cloud toward HD 24534 agrees well with 
the measured gas-phase O/H in diffuse cloud sightlines 
(\markcite{MJC98}Meyer, Jura \& Cardelli 1998).

\acknowledgments

UJS acknowledges support from the HST grant GO-06542.01-95A 
and the NASA LTSARP Grant NAG5-3539 through Villanova University.

\clearpage

\begin{figure}
\caption[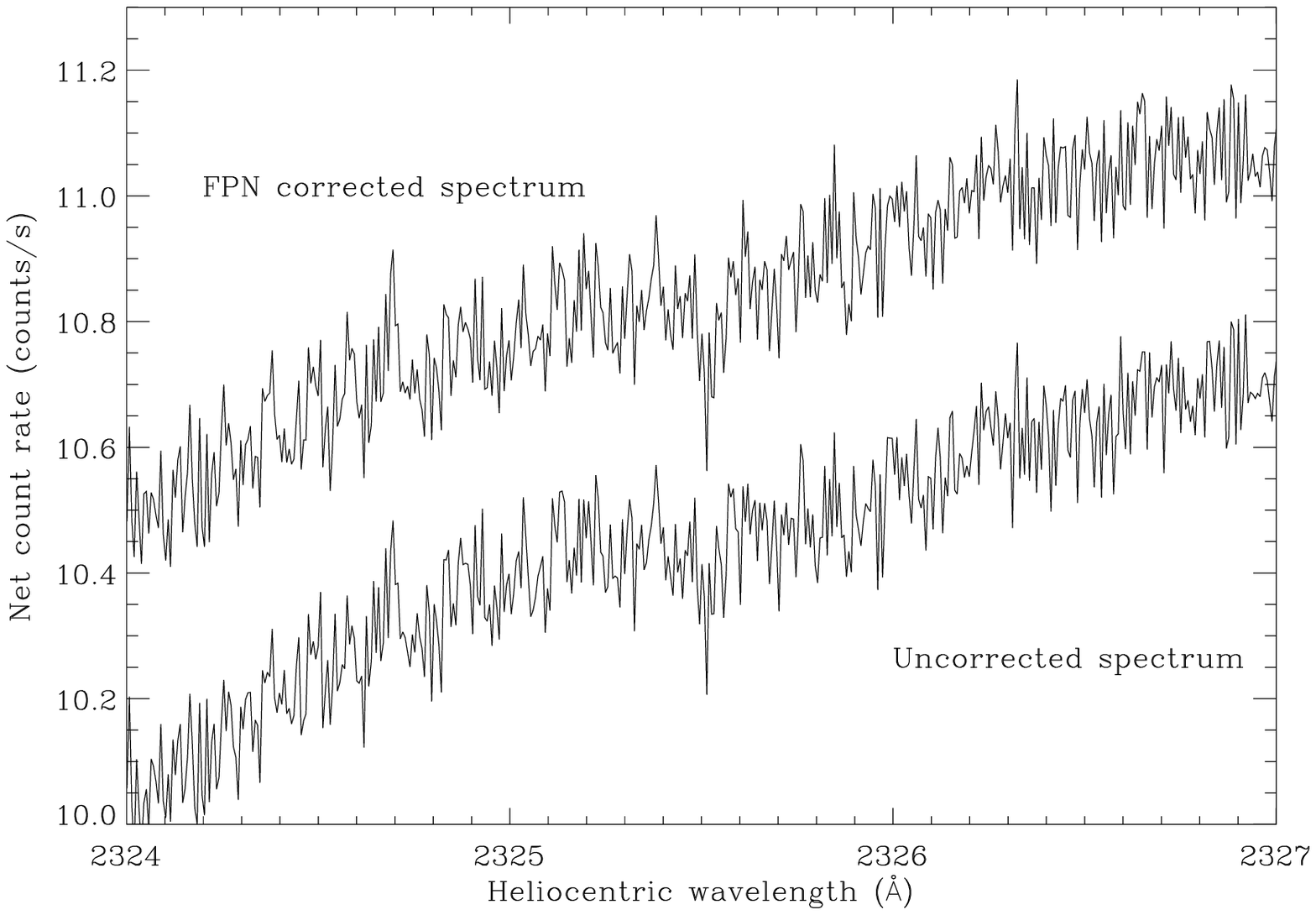]{The spectrum toward HD 24534 in the region of 
the interstellar \ion{C}{2}{\small]} 2325.403\AA\space feature. The top 
spectrum is the FPN corrected data, and the bottom spectrum is the uncorrected
data shifted by -0.4 counts/s. The \ion{C}{2}{\small]} absorption is
visible in both spectra at around 2325.5\AA.}

\caption[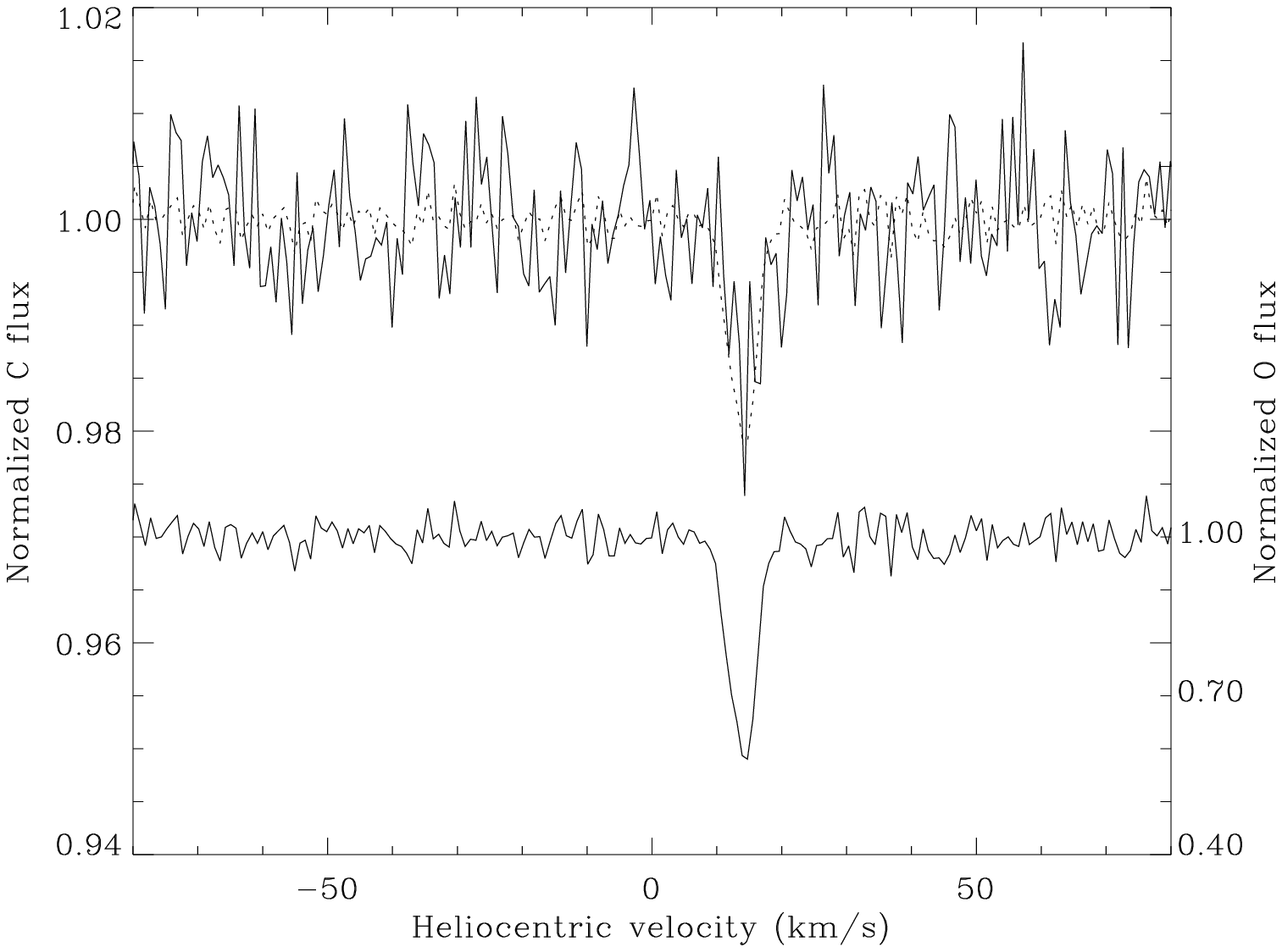]{The FPN corrected normalized spectra of the 
interstellar \ion{C}{2}{\small]} 2325\AA\space (top solid curve) 
and \ion{O}{1}{\small]} 1355\AA\space (bottom curve) features toward HD 24534.
The oxygen data have been shifted in velocity by +1.0 km/s which is well 
within the calibration uncertainty of GHRS spectra. The \ion{O}{1} data are
also shown replotted as a dotted line shifted up to the continuum level of the 
\ion{C}{2} spectrum.}

\caption[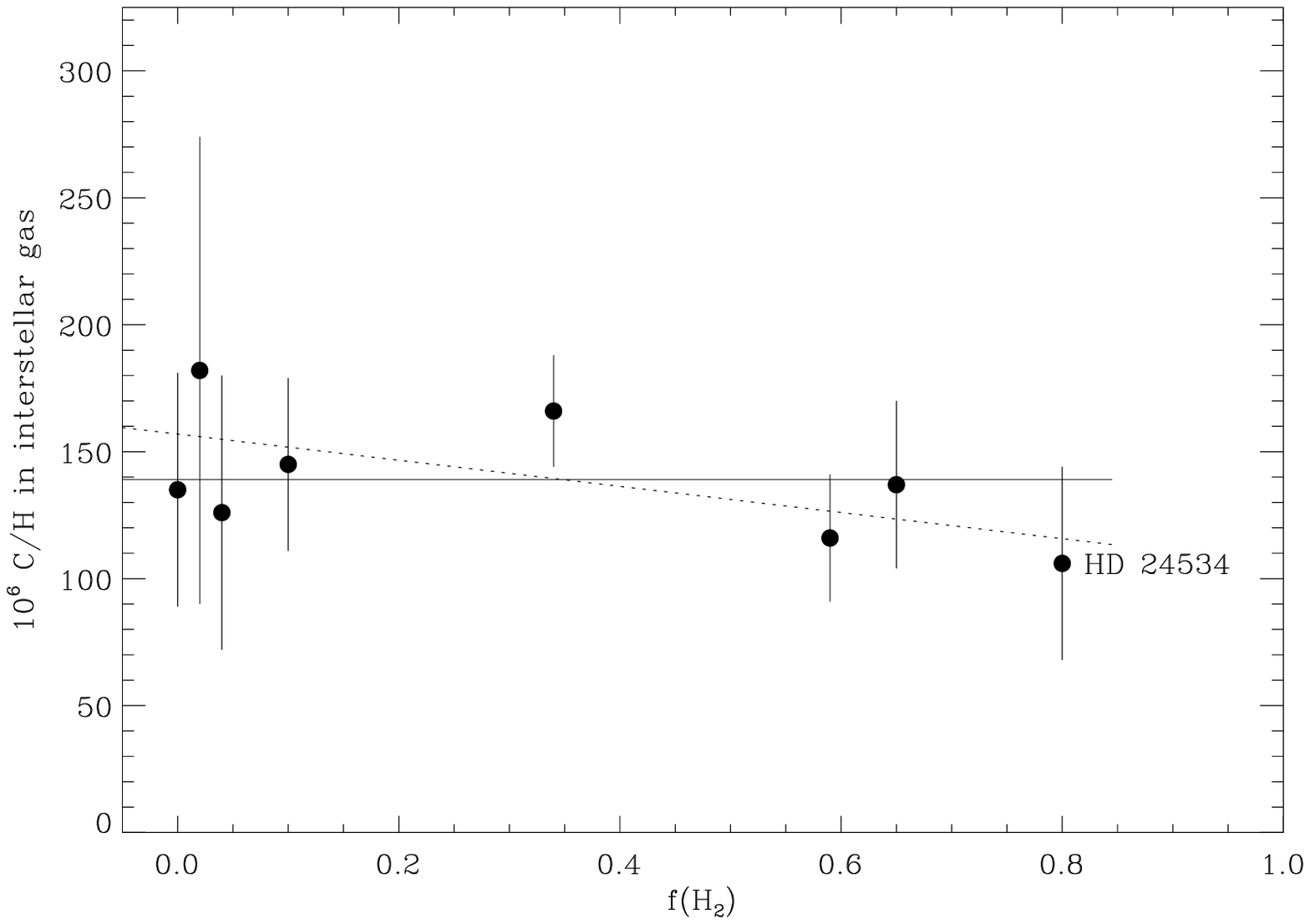]{The interstellar gas-phase carbon-to-hydrogen 
ratio multiplied by 10$^{6}$ versus the fractional H$_{2}$ abundance for all 
sightlines with weak line \ion{C}{2} measurements. The points from left to 
right represent $\tau$ CMa (Sofia et al. 1997), $\delta$ 
Sco (Hobbs, York \& Oegerle 1982), $\lambda$ Ori, $\beta^{1}$
Sco, $\xi$ Per, $\zeta$ Per, $\zeta$ Oph (Cardelli et al. 
1996) and HD 24534 (this paper). The error bars represent 
the 1$\sigma$ uncertainties of the ratios. Also shown are a line at 10$^{6}$
C/H = 139 which is the weighted average of the values, and a weighted linear 
fit to the data points (dotted line).}

\end{figure}


\clearpage

\plotone{figure1.eps}

\clearpage

\plotone{figure2.eps}

\clearpage

\plotone{figure3.eps}


\clearpage


\begin{references}

\reference{BSD78}
Bohlin, R. C., Savage, B. D., \& Drake, J. F.\ 1978, \apj, 224, 132


\reference{CARD95}
Cardelli, J. A.\ 1995, in Calibrating Hubble Space 
Telescope in the Post-Refurbishment Era, eds. A. Koratkar \& C. Leitherer
(STScI: Baltimore), p. 173


\reference{CE94}

Cardelli, J. A., \& Ebbets, D. C.\ 1994, in Calibrating Hubble Space 
Telescope, HST Calibration Workshop, ed. J.C. Blades \& 
A.J. Osmer, (STScI: Baltimore), p. 322

\reference{CARD93}
Cardelli, J. A., Mathis, J. S., Ebbets, D. C. \& Savage, B. D. 
\ 1993b, \apj, 402, L17

\reference{CARD96}
Cardelli, J. A., Meyer, D. M., Jura, M. \& Savage, B. D. 
\ 1996, \apj, 467, 334

\reference{DS94}
Diplas, A., \& Savage, B. D. 1994, ApJS, 93, 211

\reference{DRAI89}
Draine, B. T.\ 1989, in IAU Symposium 135, Interstellar Dust,
ed. L. J. Allamandola \& A. G. G. M. Tielens (Dordrecht: Kluwer)

\reference{DW81}
Duley W. W. \& Williams, D. A.\ 1981, \mnras, 196, 269

\reference{FANG93}
Fang, Z., Kwong, V. H. S., Wang, J., \& Parkinson, W. H.\ 1993,
Phys. Rev. A., 48, 1114

\reference{HOBB82}
Hobbs, L. M., York, D. G., \& Oegerle, W.\ 1982, \apj, 252, L21

\reference{JLM92}
Joblin, C., L\'eger, A., \& Martin, P.\ 1992, \apj, 393, L79

\reference{LENN85}
Lennon, D. J., Dufton, P. L., Hibbert, A., \& Kingston, A. E.\ 1985, \apj, 294,
200

\reference{MASO76}
Mason, K. O., White, N. E., Sanford, P. W., Hawkins, F. J., Drake, J. F., \& 
York, D. G.\ 1976, \mnras, 176, 193

\reference{MJC98}
Meyer, D. M., Jura, M. \& Cardelli, J. A. \ 1998, \apj, 493, 222

\reference{SCS92}
Savage, B. D., Cardelli, J. A., \& Sofia, U. J.\ 1992, \apj, 401, 706

\reference{SNOW96}
Snow, T. P., Black, J. H., van Dishoeck, E. F., Burks, G., Crutcher, R. 
M., Lutz, B. L., Hanson, M. M., \& Shuping, R. Y. \ 1996, \apj, 465, 245

\reference{SNOW98}
Snow, T. P., Hanson, M. M., Black, J. H., van Dishoeck, E. F., Crutcher, R. 
M., \& Lutz, B. L. \ 1998, \apj, 496, L113

\reference{SOFI97}
Sofia, U. J., Cardelli, J. A., Guerin, K. P. \& Meyer, D. M \ 1997, \apj, 
490, L103

\reference{SF93}
Spitzer, L. \& Fitzpatrick, E. L. \ 1993, \apj, 409, 299

\end{references}
\end{document}